\documentclass[12pt]{iopart}

\usepackage{iopams}  			  
\usepackage{graphicx}
\usepackage{subcaption}
\usepackage[american]{babel}
\usepackage{hyperref}	          
\usepackage{epstopdf}

\begin{document}

\title[Fluid dynamics in porous media with Sailfish]{Fluid dynamics in porous media with Sailfish}

\author{Rodrigo C. V. Coelho$^{1,2}$, Rodrigo F. Neumann$^{2}$ }
\address{$^1$ Instituto de F\'isica, Universidade Federal do Rio de Janeiro (UFRJ), Caixa Postal 68528, Rio de Janeiro 21941-972, Brazil}
\address{$^2$ IBM Research, Av. Pasteur 138 \& 146, Urca, Rio de Janeiro, CEP 22290-240, Brazil}
\ead{\mailto{rcvcoelho@if.ufrj.br}, \mailto{rneumann@br.ibm.com}}

\begin{abstract}
In this work we show the application of Sailfish to the study of fluid dynamics in porous media.
Sailfish is an open-source software based on the lattice-Boltzmann method.
This application of computational fluid dynamics is of particular interest to the oil and gas industry and the subject could be a starting point for an undergraduate or graduate student in physics or engineering.
We built artificial samples of porous media with different porosities and used Sailfish to simulate the fluid flow through in order to calculate permeability and tortuosity.
We also present a simple way to obtain the specific superficial area of porous media using \texttt{Python} libraries.
To contextualize these concepts, we test the Kozeny--Carman equation, discuss its validity and calculate the Kozeny's constant for our artificial samples.

\end{abstract}


\section{Introduction}

A porous medium is characterized by containing pores, i.e. void space, in its interior.
These pores can be all connected, as in a sand pack, or not, as in Styrofoam.
If the goal is to study the fluid flow in porous media, only those with connected pores must be considered.
The understanding of fluid dynamics properties of porous media is particularly relevant to the oil and gas industry, since oil is found in underground porous rocks (reservoirs).
It is crucial to estimate the permeability of a rock reservoir to hydrocarbons in order to assist decision-making and oil recovery strategies.
Permeability can be measured experimentally or calculated as a function of other porous rock properties such as porosity, tortuosity and specific surface area.
The most popular empirical relationship use to calculate the permeability from those quantities is the Kozeny--Carman equation~\cite{kozeny1927kapillare, carman1997fluid}, but there are many other formulas for more specific purposes~\cite{Xu2008}.

In this context, computational fluid dynamics (CFD) plays an important role and the Lattice Boltzmann Method (LBM) has many advantages over other methods making it a good choice to address systems with complex geometries.
LBM is a relatively new CFD technique with increasing popularity over the last twenty years.
It was very successful in simulating complex flows, such as fluids with immiscible components, interfacial problems and flows in complex geometries (e.g., in porous media).
In recent years, LBM has been extended even to semiclassical~\cite{coelho14} and relativistic fluids~\cite{mendoza10}.
The advantage over other methods lies in the simplicity of its dynamics, easy handling of complex geometries and, especially, its flexibility for implementation in parallel computing.
Sailfish~\cite{Januszewski2014} is a open-source LBM/CFD solver that comes with many examples ready to use.
It has a simple \texttt{Python} interface, which takes little to learn and is already optimized for Graphics Processing Units (GPUs).
Its advantages make it an appropriate choice for the study of flow in porous media, even for beginners in this subject.

In this work we propose a simple and efficient way to study fluid dynamics in porous media by using Sailfish.
We built porous media samples artificially by placing obstacles in random positions allowing them to overlap.
There are other models in literature to build artificial porous media.
Many authors adopt a porous medium made of identical spheres placed in a regular lattice~\cite{Pan2006,santos05}.
Koponen, Kataja and Timonen~\cite{koponen97,koponen96} used a two-dimensional porous media composed by identical squares placed in random positions.
We show how to calculate the most important fluid dynamics properties of porous media, in order to analyze the applicability of Kozeny--Carman equation to our artificial samples.
We also present an original and simple method to measure the specific surface area (SSA) from a digital rock tomography based on \texttt{Python} image-processing libraries.
All the content of this work was produced with open-source software.

This paper is organized as follows.
In Sec.~\ref{the-sailfish-sec} an introduction to Sailfish is presented.
In Sec.~\ref{artificial-porous-media} the algorithm used to build artificial samples is explained.
In Sec.~\ref{permeability-sec} we demonstrate how the permeability was calculated using the output from simulation.
In Sec.~\ref{KC-eq-section} we apply the Kozeny--Carman equation to our samples and calculate its Kozeny's constant.
In Sec.~\ref{specific-surface-area} our method to calculate the SSA is presented and tested for a simple case and, in Sec. \ref{tortusity-sec}, a simple way to calculate the tortuosity from the output data is shown.
Finally, in Sec. \ref{conclusion-sec} a discussion about the results and our concluding remarks are provided.

\section{The Sailfish}
\label{the-sailfish-sec}

This section contains a brief introduction to Sailfish~\cite{Januszewski2014}, the software we used to simulate fluid flow in our artificially-created porous media.
We do not intend to provide a tutorial, for a very detailed one can be found at the developer's website~\footnote{Download Sailfish and see the tutorial at \url{http://sailfish.us.edu.pl}}, where you can download and follow the instructions to start using it.
Sailfish requires no actual installation step, since the code is written in an interpreted programming language: \texttt{Python}.
The required packages are all enumerated in Sailfish's website.
A graphics card or GPU is also needed to run the calculations in parallel.

The main advantage of Sailfish is its ease of use.
It comes with many strategic examples that can be run and easily adapted to the needs of the user.
As all routines and examples are written in \texttt{Python}, they are easy to read and write -- as compared to compiled languages like \texttt{Fortran} and \texttt{C/C++} -- without compromising the computational performance, for all the time-consuming calculations run in the massively-parallel GPU.
Sailfish offers a high-level programming interface with several built-in functionalities, so that the user only needs to provide simple instructions like the boundary and initial conditions, physical parameters of the fluid and select the LBM relaxation dynamics.
Programming general-purpose GPUs is usually a difficult task, but Sailfish hides this complexity away and makes all parallelization behind the scenes~\cite{Januszewski2014}.
Simulations in Sailfish can be made interactive which, combined to its high performance, makes it a powerful tool to illustrate hydrodynamic principles in the classroom in real-time.
While a simulation is running in visualization mode, one can add new (solid) obstacles by freehand drawing them with a cursor.

\begin{figure}[htb]
\center
\includegraphics[width=0.5\linewidth]{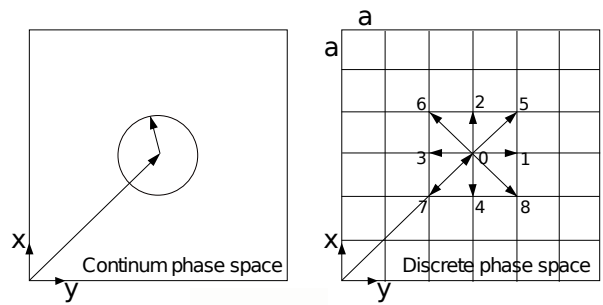}
\caption{Discretization of position and velocity spaces using the D2Q9 lattice.}
\label{discrete-space}
\end{figure}

The LBM~\cite{mohamad}, in which Sailfish is based, is a numerical method to solve the Boltzmann equation~\cite{kremer}, that in its finite-difference form is
\begin{equation}
\frac{f(\mathbf{x}+\boldsymbol{\xi}\Delta t, \boldsymbol{\xi} + \frac{\mathbf{F}}{m} \Delta t , t + \Delta t) - f(\mathbf{x}, \boldsymbol{\xi}, t)}{\Delta t} = \left(\frac{\partial f}{\partial t}\right)_{\mathrm{coll}}, 
\label{EqBoltzmann}
\end{equation}
where $\mathbf{x}$ is the position, $\boldsymbol{\xi}$ is the microscopic velocity, $t$ is the time, $\mathbf{F}$ is the external force and $m$ is mass of the particles.
The above equation gives the time evolution of the distribution function, $f(\mathbf{x}, \boldsymbol{\xi}, t)$, by knowing the collision operator, $\left(\partial f/\partial t\right)_{\mathrm{coll}}$, which includes all information about atomic aspects of scattering process of the particles in the gas. 
The Bhatnagar-Gross-Krook (BGK) collision term~\cite{bhatnagar1954model},
\begin{equation*}
\left(\frac{\partial f}{\partial t}\right)_{\mathrm{coll}} = - \frac{f(\mathbf{x}, \boldsymbol{\xi}, t) - f^{\mathrm{eq}}(\mathbf{x}, \boldsymbol{\xi}, t)}{\tau},
\end{equation*}
is the simplest and most commonly used in LBM. 
It assumes that the non-equilibrium function $f(\mathbf{x}, \boldsymbol{\xi}, t)$ tends to the equilibrium function $f^{\mathrm{eq}}(\mathbf{x}, \boldsymbol{\xi}, t)$ with a characteristic time $\tau$, called ``relaxation time''. 
In LBM the position and velocity spaces are discretized as illustrated in Fig. \ref{discrete-space}. 
The time step $\Delta t$, used in \eref{EqBoltzmann}, is the time for a particle move from its site to the first neighbor site and this time step with the lattice parameter ($a$) form a system of unities called \textit{lattice units}~\cite{mohamad}, which can be converted for any other system of unities.

\section{Fluid flow in porous media}
\label{basic-comcepts-sec}

The most relevant quantities related to fluid dynamics in porous media are porosity, tortuosity, surface area and permeability.
In this section we define these concepts and relate them using the Kozeny--Carman equation.

We start with the porosity, that is defined as the fraction $\phi = V_{\mathrm{pores}} / V_{\mathrm{total}}$ of the total volume that is occupied by connected pores.
Non-connected pores, as in Styrofoam, do not allow flow and, therefore, are out of our scope.
In case the medium has both connected and non-connected pores, the volume occupied by the latter must be disregarded.

\begin{figure}[htb]
\center
\includegraphics[width=0.5\linewidth]{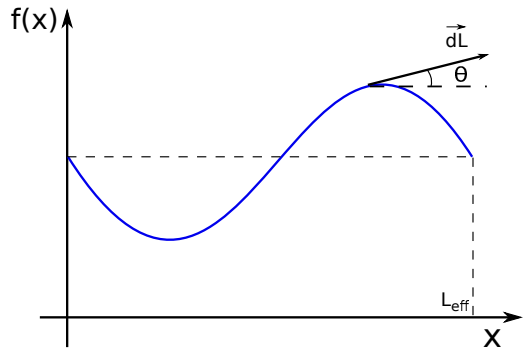}
\caption{Example of a fluid streamline for tortuosity calculations.}
\label{tort-ex}
\end{figure}
Tortuosity~\cite{Clennell97, sheidegger} is a geometrical figure-of-merit that indicates how much the fluid flow streamlines deviate from straight paths.
It is defined as the ratio between the total length along the streamline and its effective length, as depicted in figure~\ref{tort-ex}.
Note that $\tau \geq 1$.
If the streamline is given by a function $f(x)$, as exemplified in figure ~\ref{tort-ex}, the tortuosity can be calculated as
\begin{equation}
\tau = \frac{\int\!dL}{L_{\mathrm{eff}}} = \frac{\int_0^{L_{\mathrm{eff}}} \sqrt{1+\left(\frac{\partial f}{\partial x}\right)^2}dx}{L_{\mathrm{eff}}}.
\label{tort-eq}
\end{equation}

The internal area of obstacles in contact with the fluid greatly influences the fluid dynamics.
Ultimately, the SSA (or ``surface-to-volume ratio'') is responsible for the strength of surface interaction effects.
This figure-of-merit is defined as $s = A_{\mathrm{obs}} / V_{\mathrm{obs}}$, where $A_{\mathrm{obs}}$ is the surface area of obstacles and $V_{\mathrm{obs}}$ is their total volume.

In general, permeability is the quantity of greatest interest in the study of flow through porous media.
It quantifies how easily a specific fluid passes through the media, i.e, the inverse of a ``flow resistance''.
In its simplest formulation, it depends only on the geometry of the medium and the viscosity of the fluid, $\mu$, and can be calculated using Darcy's law~\cite{darcy1856determination, whitaker1986flow},
\begin{equation}
\kappa _{ij} = \frac{\mu \langle u^i \rangle}{\frac{dP}{dx^j}} \, ,
\label{darcy-eq}
\end{equation}
where $\frac{dP}{dx^j}$ is the pressure gradient along direction $x^j$ and $\langle u^i \rangle$ is the average velocity of the fluid in the direction $x^i$.
Note that, in general, permeability is a $3 \times 3$ tensor, but in practical cases we are usually interested only in the diagonal components $\{ \kappa_{xx}, \kappa_{yy}, \kappa_{zz} \}$.
The SI unit for permeability is $m^2$, however, the ``darcy'' $(1\,\mathrm{darcy} = 10^{-12}\,m^2)$ is the most used.

The four concepts presented previously are related by the Kozeny--Carman equation
\begin{equation}
\kappa = \frac{1}{C s^2}\frac{\phi^3}{(1-\phi)^2},
\label{kcequation}
\end{equation}
where $C$ is an empirical and dimensionless constant, known as ``Kozeny's constant''.
Consider, for instance, a medium composed of identical spheres with diameter $d$, equally distributed in a bed packed~\cite{kaviany}.
In this case, the SSA is $s=A_{\mathrm{sphere}}/ V_{\mathrm{sphere}} = 6/d$.
So ~\eref{kcequation} becomes
\begin{equation}
\kappa = \frac{d^2}{36 C}\frac{\phi^3}{(1-\phi)^2}.
\label{kc-spheres}
\end{equation}
Note that the dimensionality is entirely contained within $d^2$, which agrees with the fact that the dimension of $\kappa$ is \textit{length-squared}.
The Kozeny--Carman is not the only existing permeability-porosity relationship~\cite{Xu2008}, but it works fairly well for granular beds as in our case.

\subsection{Artificial porous media}
\label{artificial-porous-media}

In this section we describe the algorithm we developed to build artificial samples.
An object-oriented implementation example for such algorithm, written in \texttt{Python}, is available as a Supplementary Material.
Our intention was to produce a more realistic porous medium than the usual bed packed, but with more controllable parameters than the real samples digitized from rock tomography.
In this way, we can, for instance, relate the geometry of the obstacles to permeability.

The input parameters for the algorithm are: the length of obstacles, the length of the sample and the target porosity.
The obstacle's shape is determined by the equation of a spheroid with radii $\{ r_x, r_y, r_z \}$ centered at $(x_c, y_c, z_c)$.
The samples are rectangular parallelepipeds, for which three lengths $\{ L_x, L_y, L_z \}$ are required as inputs.
The last input is the target porosity $(\phi_{\mathrm{t}})$, that is the desired porosity for the sample within a tolerance $\varepsilon$.

\begin{figure}[htb]
\center
\includegraphics[width=0.8\linewidth]{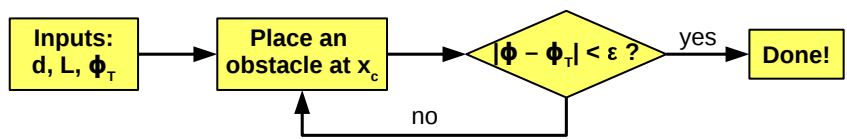}
\caption{Flowchart of the algorithm used to generate artificial porous media samples.}
\label{artificial-fluxogram}
\end{figure}

Given these input parameters, the algorithm builds a sample by placing obstacles in random positions, one at a time.
After creating an obstacle, the algorithm checks if the target porosity has been reached.
If not, another obstacle is added.
The algorithm stops when $|\phi-\phi_{\mathrm{t}}| \leq \varepsilon$, as depicted in figure ~\ref{artificial-fluxogram}.

\begin{figure}[htb]
\center
\includegraphics[width=\linewidth]{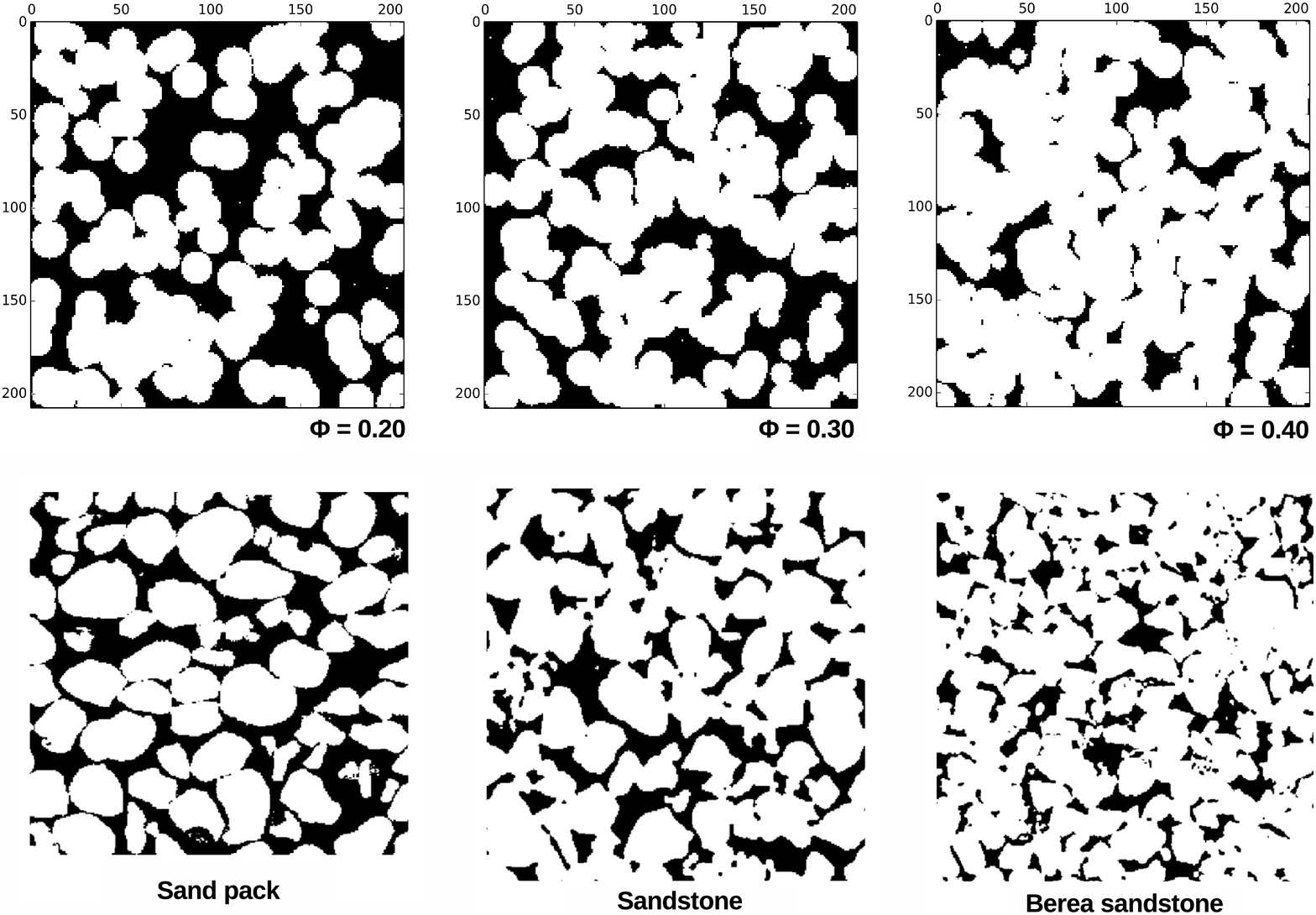}
\caption{Comparison between slices of our artificial samples and of real rocks taken from the literature~\cite{rabbani14}.
The artificial media were built using spherical obstacles with diameter of 20 nodes and with target porosity as indicated in each figure.}
\label{artificial-blocs-por}
\end{figure}

For our simulations, we used samples with $208^3$ nodes, spherical obstacles with diameters between $d = 12$ and $d = 30$ and porosities in the range $0.2 \leq \phi \leq 0.4$.
The number of nodes $(\sim 9 \times 10^6)$ was limited by the available GPU memory.
In figure ~\ref{artificial-blocs-por}, we see some slices of artificial samples generated by our algorithm side-by-side with slices of real digitized rock tomographies for comparison.
These artificial samples have the same obstacle diameter $(d=20)$ but different porosities: $0.2$, $0.3$ and $0.4$.
Since the spheres are allowed to overlap, they form more complex structures that are very close to those observed in real porous media.
This leads to more realistic simulations of fluid flow.

\subsection{Permeability}
\label{permeability-sec}

The two main methods for inducing fluid flow in a porous medium using Sailfish are: by setting a pressure difference across the sample or a body force acting on every fluid node.
Although the pressure method is closer to what is done in experiments, we observed that equilibrium is reached much faster with the force method.
This happens because all points of the fluid feel the driving force at the same time when using the force method, instead of by propagation of shock waves as in the pressure method.
In order to detect when equilibrium is achieved in our simulations, we calculate $\left \vert \vert\boldsymbol{u}_{\mathrm{new}}\vert - \vert\boldsymbol{u}_{\mathrm{old}}\vert \right \vert /\vert\boldsymbol{u}_{\mathrm{new}}\vert$ at every time step for every point of the fluid and, when the maximum value of this quantity is less than a given threshold, the simulation stops.
We found $5 \times 10^{-7}$ to be sufficient for our purposes.

The permeability is calculated using Darcy's law but, as it assumes a pressure gradient instead of a body force, we modify it using the relation $\frac{dP}{dx^j} dx^j = F_j/A_j$, where $F_j$ is the body force along direction $x^j$ and $A_j$ is the cross-sectional area of the fluid at the boundary where the pressure is applied.
The cross-sectional area of the fluid is $\phi$ times the total area.
Therefore, given a force density $\mathbf{f} = f_x \mathbf{i}$, Darcy's law becomes
\begin{equation}
\kappa = \frac{\mu v_x \phi}{f_x},
\label{darcy-force}
\end{equation}
which is the equation use to calculate permeability.

\begin{figure}[htb]
\center
\includegraphics[width=0.5\linewidth]{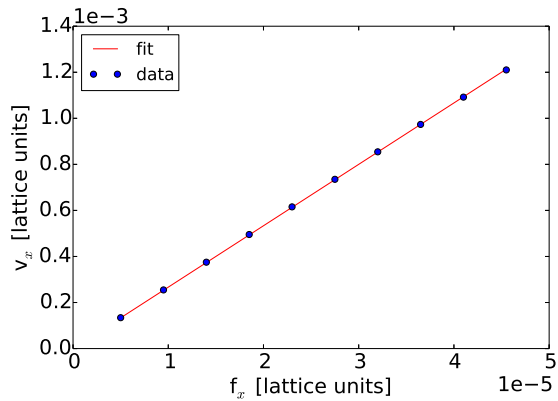}
\caption{Proportionality between the mean velocity and the applied force for a given artificial sample with $\phi =0.30$ and $d=16$.}
\label{VxF}
\end{figure}

Since permeability, porosity and viscosity are not velocity-dependent, we expect the applied force $f_x$ and the mean velocity $v_x = \langle u^x \rangle$ to be proportional to each other.
We tested this by taking an artificial sample with porosity $\phi = 0.3$, obstacles with diameters $d = 16$ nodes and using it as input for fluid flow simulations with different driving forces.
The fluid had viscosity $\mu=0.01$ in lattice units.
The result is shown is figure ~\ref{VxF}.
We see that the function $v_x = \alpha f_x$ fits well the simulated data points, with a proportionality constant of $\alpha = (26.67 \pm 0.02)$ in lattice units.
Using ~\eref{darcy-force}, we determined the permeability of the sample to be $\kappa = (8.001 \pm 0.006) \times 10^{-2}$ in lattice units.

\subsection{The Kozeny--Carman equation}
\label{KC-eq-section}

\begin{figure}[htb]
\centering
\begin{subfigure}[h]{0.49\textwidth}
  \caption{\hfill~}
  \includegraphics[width=\columnwidth]{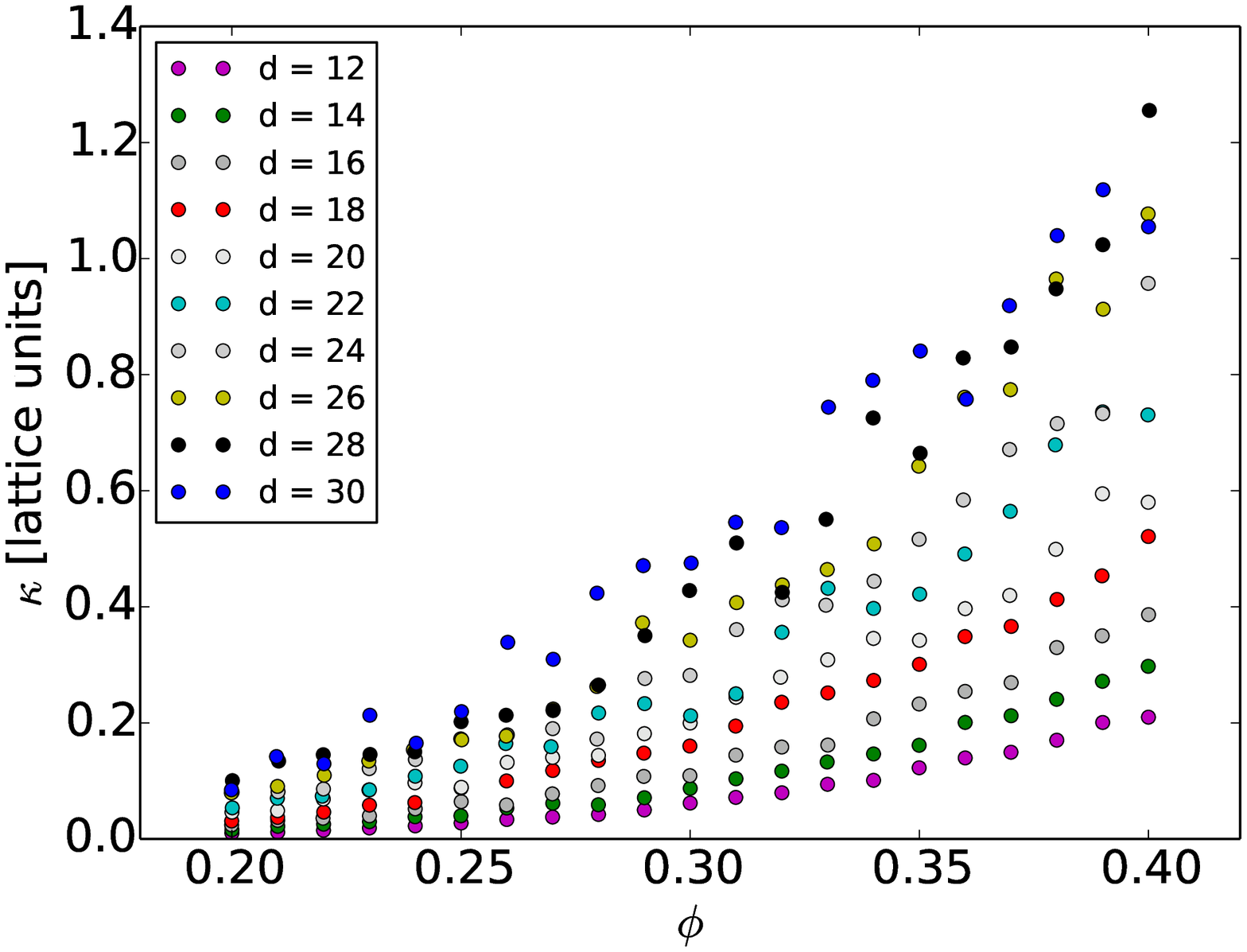}
  \label{perm-fit}
\end{subfigure}
\begin{subfigure}[h]{0.49\textwidth}
  \caption{\hfill~}
  \includegraphics[width=\columnwidth]{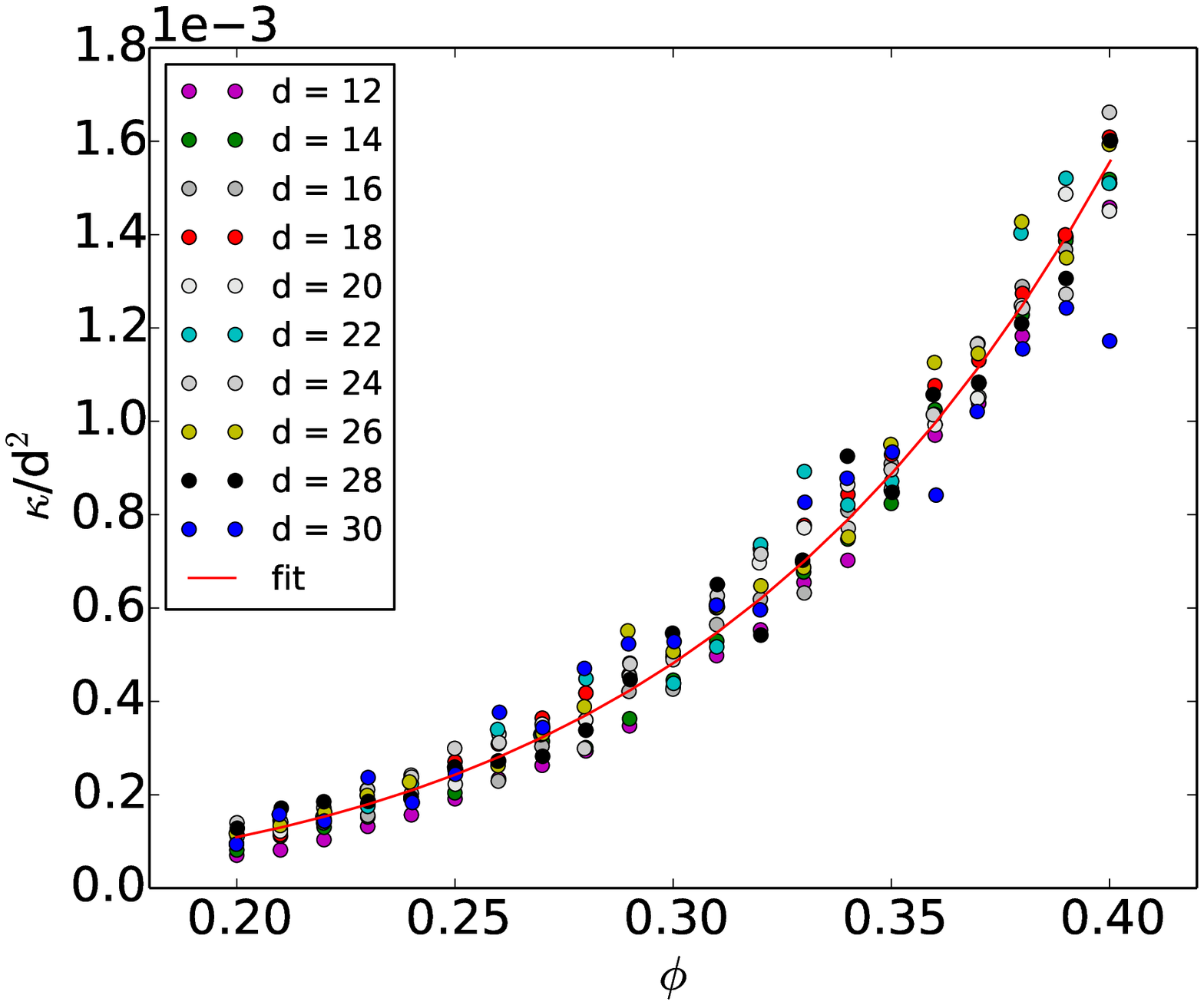}
  \label{perm-fit-scaled}
\end{subfigure}
\caption{Permeability dependence on porosity and obstacle diameter.
(\subref{perm-fit}) Permeability as a function of porosity for several obstacle diameters.
(\subref{perm-fit-scaled}) Non-dimensional permeability as a function of porosity for several obstacle diameters.
The curve fit is given by ~\eref{KC-simp} with $C =  3.18 \pm 0.02$. }
\label{fig:permeability}
\end{figure}

We calculated the permeability using ~\eref{darcy-force} and estimated the Kozeny's constant of the artificial porous media samples created using the method described in Section~\ref{artificial-porous-media}.
In other to determine the individual contributions from porosity and obstacle dimension to the permeability, we built 210 samples with $208^3$ nodes: 10 obstacle diameters $d \in [12, 30]$ and 21 porosities $\phi \in [0.2, 0.4]$, for each diameter.
The fluid viscosity was $0.01$ and the external force, $f_x = 10^{-5}$.
The results are shown in figure ~\ref{perm-fit}.
One sees that permeability increases with porosity for a given obstacle diameter.
Analogously, for a given porosity, permeability also increases with obstacle diameter.
Next, we investigate the suitability of the Kozeny--Carman equation to describe our data.

Equation~\eref{kc-spheres} tells us that permeability is proportional to $d^2$ or, in other words, that there is a scale-law involving obstacle dimensions and permeability.
In order to turn permeability into a scale-free non-dimensional quantity, we calculate $\kappa/d^2$, and confirm that all points fall (despite fluctiations) within the same curve (see figure ~\ref{perm-fit-scaled}).
We fit the function
\begin{equation}
\frac{\kappa}{d^2} = \frac{1}{36 C}\frac{\phi^3}{(1-\phi)^2}
\label{KC-simp}
\end{equation}
to the data points, with $C$ (Kozeny's constant) as the free parameter.
Its value was determined to be $C =  3.18 \pm 0.02$.
As the fit uncertainty represents less than $1\%$ of the main value, we consider ~\eref{KC-simp} to be in good agreement with our simulated data points.

For simplicity, some approximations were made while fitting ~\eref{KC-simp} to the data points.
The obstacle diameter $d$ is taken as the \textit{theoretical} diameter but the \textit{effective} one might be slightly different due to the ``digital'' nature of the spheres, i.e., being composed of cubic pixels (see figure ~\ref{sphere-pixels}).
Also, we used the SSA of non-overlapping spheres, $s=6/d$, although we allow the spheres to overlap.
In the next sections we present an analysis of the impact of such approximations.

Despite those simplifications, the value we obtained for Kozeny's constant is consistent with previously published values in the literature.
For instance, $C=5$ for a bed packed of non-overlapping spheres~\cite{Xu2008} and, typically, between $2.2$ and $8.9$ for a fibrous media~\cite{Yazdchi2011}.

\subsection{Specific superficial area}
\label{specific-surface-area}

With the increasing popularization of Digital Rock Physics~\cite{andra2013digital-I, andra2013digital-II}, some recent works proposed new techniques to calculate the SSA~\cite{rabbani14}, but they are still quite complex and indirect.
Measuring the SSA of porous rocks experimentally is a hard task~\cite{lowell12} but it is important to understand the rock's properties~\cite{Suvachittanont96}.
We propose a methodology to calculate the SSA using widely-available \texttt{Python} libraries for image-processing.
The advantage of our method is that it provides a visual interpretation of what is being calculated, making it simple enough to be understood and applied by beginners.

To calculate the SSA of a digital rock sample, we use the ``binary\_erosion'' method of the SciPy library~\footnote{See \url{http://www.scipy-lectures.org/advanced/image_processing/}}.
This operation \textit{erodes} the image by removing a layer of pixels at the rock/pore boundary.
When the eroded image is subtracted from the original, what remains are the pixels at the surface, which allows us to calculate the SSA using its basic definition.
A \texttt{Python} implementation example is provided in the Supplementary Material section.

\begin{figure}[htb]
\centering
\begin{subfigure}[h]{0.5\textwidth}
  \caption{\hfill~}
  \includegraphics[width=\columnwidth]{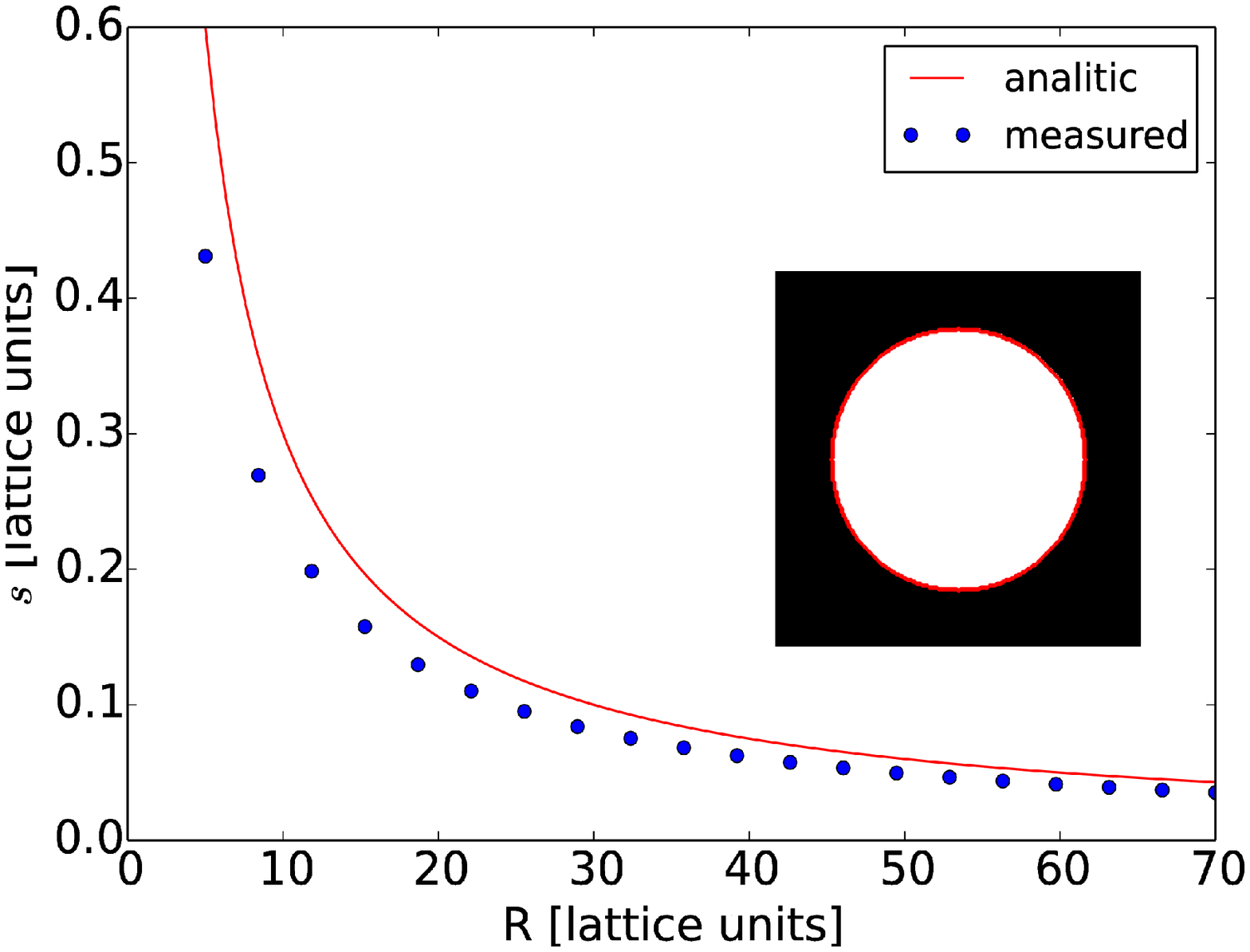}
  \label{area-sphere}
\end{subfigure}
\begin{subfigure}[h]{0.6\textwidth}
  \caption{\hfill~}
  \includegraphics[width=\columnwidth]{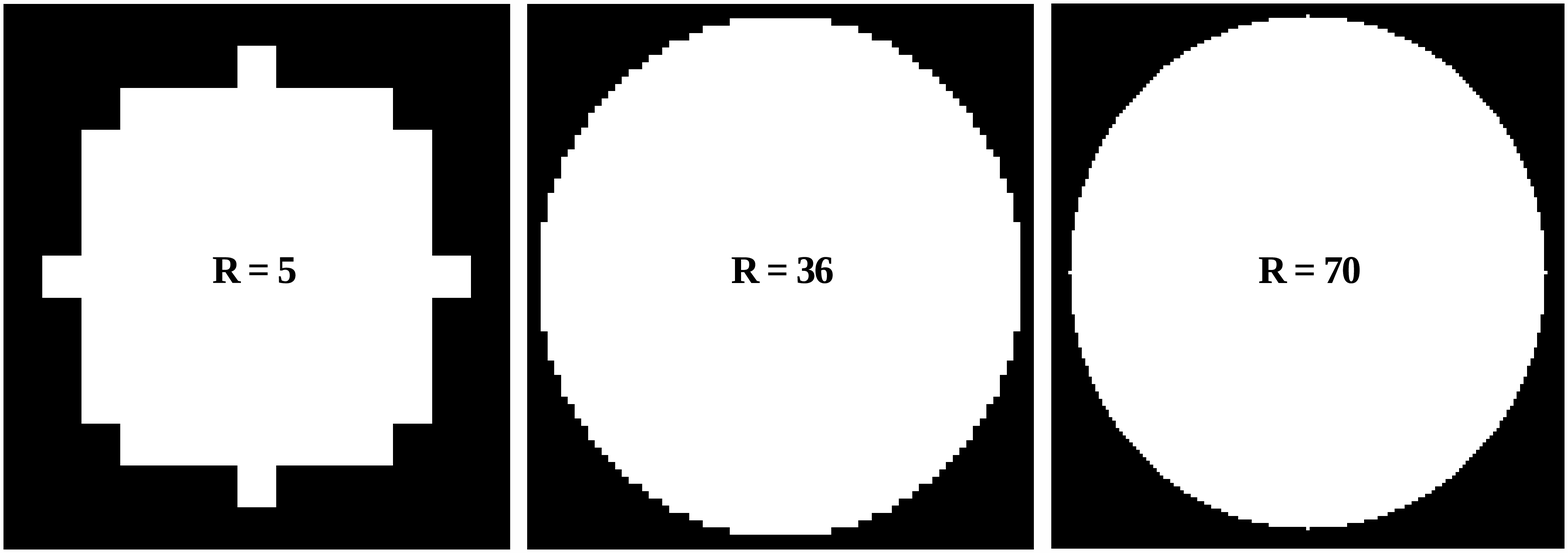}
  \label{sphere-pixels}
\end{subfigure}
\caption{Specific surface area calculated according to the method in Sec.~\ref{specific-surface-area}.
(\subref{area-sphere}) Specific surface area for a single sphere as a function of radius, compared to the analytical expression.
The inset shows a slice of a sphere with $R=70$ and its surface area in red.
(\subref{sphere-pixels}) Slices at the equator of digital spheres with different radii.
The larger the radii, the closer the figure gets to a spherical shape.}
\label{fig:SSA1}
\end{figure}

The first application of this method is a simple problem with analytical solution that consists in calculating the SSA of a single sphere for different radii.
In figure ~\ref{area-sphere}, we see the results for radii ranging from $5$ to $70$ compared to the analytical solution, $s=3/R$.
We see that the data follows the same behavior of the analytical curve, but with a small discrepancy that vanishes as the radius increases.
This difference is due to a limitation in the spatial resolution that all digital images are subjected to.
In figure ~\ref{sphere-pixels} we see three slices of spheres with different radii.
The sphere with $R=5$ is far from a spheroidal shape, while that with $R=70$ is much closer.
The better the spatial resolution, the more accurate is the proposed methodology.
However, since the available GPU memory is limited, we cannot use arbitrarily large sample sizes to achieve higher resolutions.

\begin{figure}[htb]
\centering
\begin{subfigure}[h]{0.49\textwidth}
  \caption{\hfill~}
  \includegraphics[width=\columnwidth]{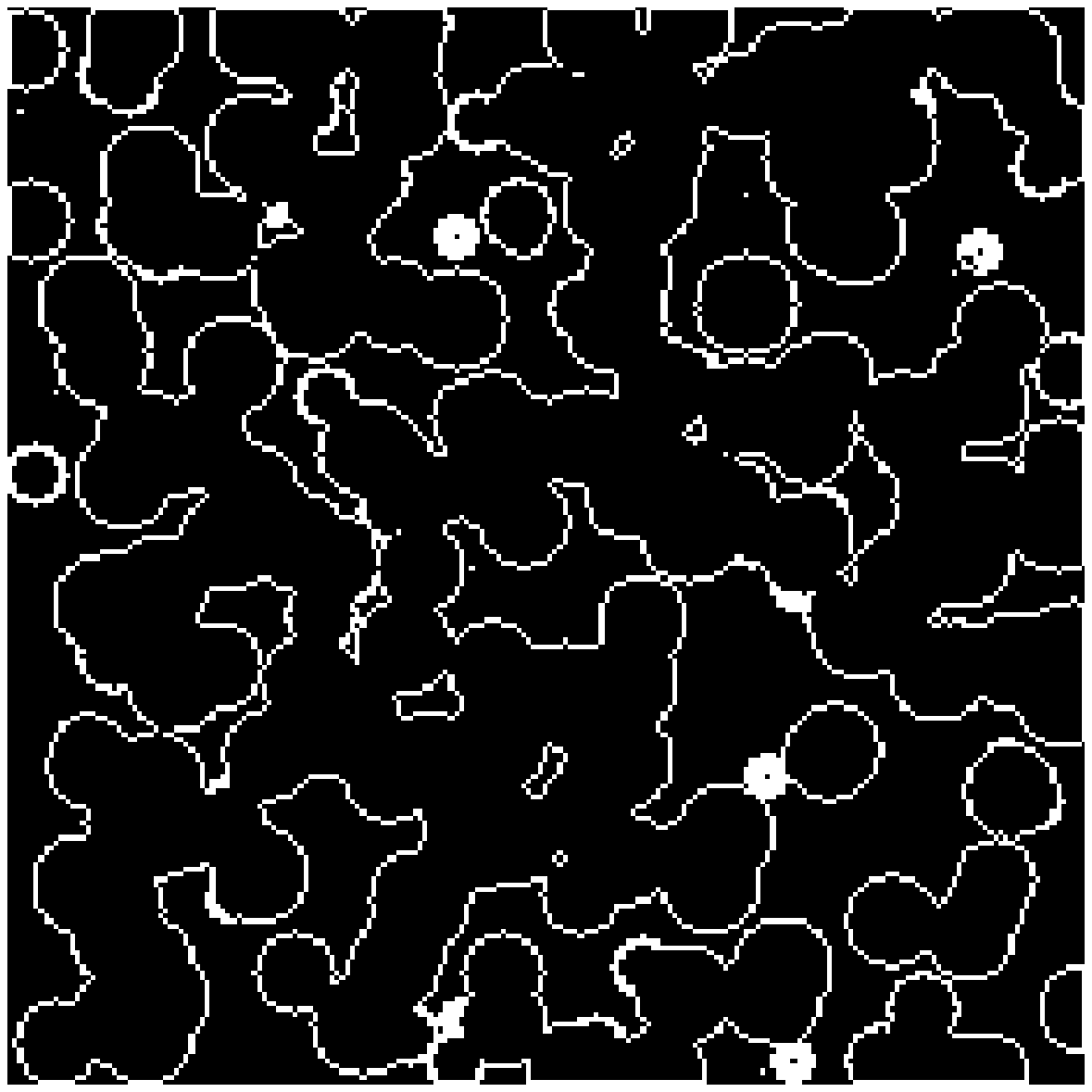}
  \label{surface0_30}
\end{subfigure}
\begin{subfigure}[h]{0.49\textwidth}
  \caption{\hfill~}
  \includegraphics[width=\columnwidth]{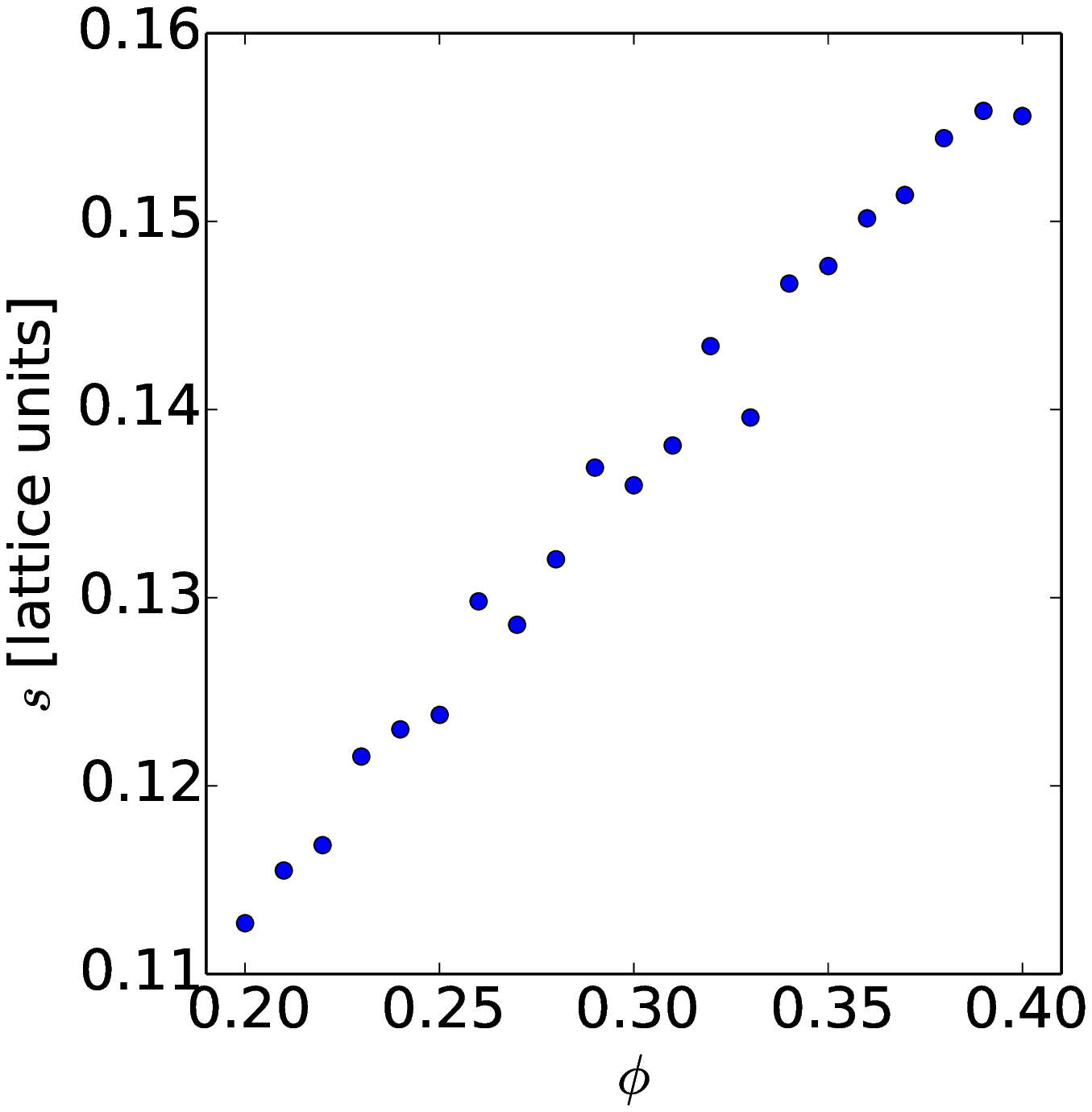}
  \label{area-por}
\end{subfigure}
\caption{Specific surface are for porous media built using $d=20$ nodes as obstacle diameter.
(\subref{surface0_30}) A slice showing the SSA of an artificial porous media sample with $\phi=0.3$.
(\subref{area-por}) Specific surface area as a function of porosity for a given obstacle diameter.}
\label{fig:SSA2}
\end{figure}

After this simple, but revealing, example we calculated the SSA for our porous media samples.
figure~\ref{surface0_30} shows the surface of the obstacles, that is, the outermost layer isolated using the erosion operation.
We clearly see the contour of the overlapped spheres and some partially filled circles, due to the finite resolution issue.
In figure ~\ref{area-por} we show the SSA obtained as a result of our method for 21 samples with $d=20$ and different porosities.
The SSA clearly depends on porosity, which was not considered in our previous calculation of the Kozeny's constant.
This dependence occurs due to obstacles being able to overlap~\cite{koponen97}, which changes their shape to non-spherical.
If the spheres were isolated from each other, as in a close-packed structure, the SSA would be $s=3/R$ as for a single sphere.

\subsection{Tortuosity}
\label{tortusity-sec}

In general, is difficult to calculate tortuosity using its basic definition, ~\eref{tort-eq}, for one would have to know exactly the central streamline.
However, if one has the velocity field, there is an easier way to perform this calculation.
It has been shown~\cite{koponen96} that ~\eref{tort-eq} is equivalent to
\begin{equation}
\tau = \frac{\langle | \boldsymbol{u} | \rangle}{\langle u^j \rangle},
\label{tort-method}
\end{equation}
where $|\boldsymbol{u}|$ is the modulus of local velocity, $u^j$ is the local velocity along the direction of flow $x^j$ and $\langle \cdot \rangle$ represents a spatial average.
This method has been extensively tested~\cite{matyka12} and its results agree with theoretical predictions.

\begin{figure}[htb]
\center
\includegraphics[width=0.5\linewidth]{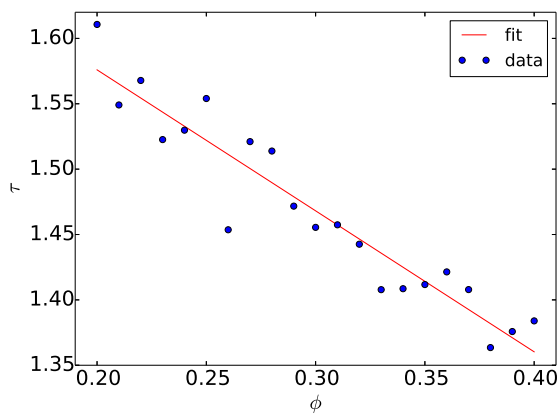}
\caption{Relationship between tortuosity and porosity for samples with obstacle diameter $d=20$.}
\label{tort-por}
\end{figure}

The tortuosity was calculated using the aforementioned method for samples with $d=20$, as in figure ~\ref{tort-por}.
We found a linear relationship between $\tau$ and $\phi$ in this range of porosities, despite all the noise, similar to previous reports in the literature~\cite{koponen96,koponen97}.
We adjusted a linear function to the data points and obtained $\tau(\phi) = 1.79 - 1.08 \phi$.

\section{Conclusion}
\label{conclusion-sec}

This present article had two goals: (i) constructing an algorithm to build artificial porous media samples and (ii) presenting how the main properties of porous media are calculate using the output from LBM simulations.
The first goal was addressed in Sec.~\ref{artificial-porous-media}.
Identical spheroidal obstacles were placed in random positions (allowing overlap), creating structures similar to those observed in real rock samples, as exemplified in figure~\ref{artificial-porous-media}.

To achieve the second goal, we developed the following analysis.
In Sec.~\ref{permeability-sec} we calculated the permeability of a particular sample using Darcy's law and verified that the drift velocity is linear to the external force.
In Sec.~\ref{KC-eq-section} we investigated whether the Kozeny--Carman equation applies to the permeability-porosity relationship of our samples, obtaining a good agreement as seen in figure ~\ref{perm-fit-scaled}.
Although we allowed obstacles to overlap in our artificial samples, destroying their otherwise spherical shape, figure ~\ref{perm-fit} shows that permeability still depends on the obstacle diameter, which allowed us to use a scale-law leading to figure ~\ref{perm-fit-scaled}.
In Sec.~\ref{specific-surface-area}, we developed our technique for calculating the specific surface area in digital samples, tested it on a single sphere for different radii and obtained good agreement to the analytic solution for big radii.
In Sec.~\ref{tortusity-sec} we presented a simple method to calculate tortuosity from the velocity field of the fluid.

Sailfish is a powerful LBM solver with many features that are out of the scope of this work.
More complex flows can be simulated in porous media using Sailfish, such as multiphase~\cite{Huang11} and turbulent flows~\cite{Ren16}.
One can also use Sailfish and the techniques described here to perform fluid dynamics studies inside fractures, which is another interesting subject for the oil and gas industry.
The relationship between SSA and porosity in porous media is yet another interesting subject for future works~\cite{koponen97}.

\ack
R.C.V. Coelho thanks CNPq for financial support.

\section*{References}
\bibliography{reference}

\end{document}